\documentclass[journal]{IEEEtran}

\usepackage{cite}
\usepackage{amsmath,amssymb,amsfonts}
\usepackage{amsthm}
\usepackage{graphicx}
\usepackage{textcomp}
\usepackage{enumitem}
\usepackage{booktabs}
\usepackage{tabularx}
\usepackage{threeparttablex}
\usepackage{algorithm}
\usepackage{algpseudocode}
\usepackage{tikz}
\usetikzlibrary{arrows.meta,positioning,fit,shapes.geometric}
\usepackage{pgfplots}
\usepackage{soul,xcolor}
\usepackage{adjustbox}
\usepackage[hidelinks]{hyperref}

\pgfplotsset{compat=1.18}

\def\BibTeX{{\rm B\kern-.05em{\sc i\kern-.025em b}\kern-.08em
T\kern-.1667em\lower.7ex\hbox{E}\kern-.125emX}}

\begin{document}

\title{Efficient and Implementation-Hardened RBLWE on Commodity Cortex-M Microcontrollers}
\author{Muhammad Asim Javaid, Muhammad Adeel Pasha, and Muhammad Ali Siddiqi
\thanks{The authors are with the Department of Electrical Engineering, Lahore University of Management Sciences, Lahore, Pakistan.}}

\maketitle

\begin{abstract}
Efficient post-quantum cryptography on resource-constrained Internet-of-Things (IoT) devices requires implementations that exploit the target processor architecture while resisting practical implementation attacks. This paper presents an ISA-accelerated and implementation-hardened realization of Ring Binary Learning with Errors (RBLWE) encryption on a commodity ARM Cortex-M33 microcontroller. Packing four 8-bit polynomial coefficients into the byte lanes of a 32-bit register and processing them with SIMD-style instructions, together with a packed message codec, accelerates encryption and decryption, while a buffered hardware-TRNG entropy source drawn from the on-die Secure Element drives the key-generation gain. Together these give same-core cold-start speedups of $4.06\times$, $3.35\times$, and $3.01\times$ for key generation, encryption, and decryption over a scalar baseline, and $3.52\times$/$3.18\times$ lower encryption/decryption cycle counts than a reference Cortex-M0 implementation. On top of this accelerated core, we add \textit{four} staged countermeasures, \textit{constant-time execution}, \textit{fault hardening} against zeroing, random-corruption, and instruction-skip faults, a \textit{Fujisaki--Okamoto (FO)-style CCA2 transform}, and \textit{first-order shared (masked) CPA decryption}, reporting each layer's cost individually. Binary-level inspection confirms these countermeasures survive compilation, and identifies a compiler-induced \textit{masking flaw} resolved with a hand-written assembly replacement. Dudect-style timing tests, debugger-assisted fault-injection campaigns, and component-level TVLA then provide implementation-level evidence for the staged protections. The results demonstrate a practical acceleration-security tradeoff for RBLWE on off-the-shelf microcontrollers and reusable architecture-aware techniques for lightweight post-quantum implementations.
\end{abstract}


\section{Introduction}
The emergence of quantum computing threatens the long-term security of conventional public-key cryptography, motivating the shift to Post-Quantum Cryptography (PQC) \cite{buchanan2017will}, with lattice-based schemes such as ML-KEM occupying a central position \cite{avanzi2019crystals,alagic2022status}. However, for embedded and IoT deployments, algorithmic security is not sufficient: implementations must meet strict latency, memory, and energy budgets while resisting physical attacks on devices that are inexpensive, battery-powered, and accessible to an adversary~\cite{fernandez2019pre}. Ring Binary Learning with Errors (RBLWE)~\cite{buchmann2016high} is attractive for exactly this setting: by drawing binary rather than Gaussian secrets and errors, it reduces sampling and arithmetic complexity relative to general Ring-LWE~\cite{lyubashevsky2013ideal,khalid2021axrlwe}, making it a natural candidate for IoT nodes with the smallest footprint.

RBLWE acceleration remains an active research goal, evidenced by a steady stream of recent FPGA and dedicated-hardware designs~\cite{he2023fpga,he2023kina,bao2023tina,ahmadunnisa2024x2dl,ahmadunnisa2024cnc,he2025high,bao2025chirp} that report substantial throughput and area-time product gains. These designs, however, accelerate the arithmetic in isolation: they report cycle counts or area-time products without evaluating the design against side-channel or fault adversaries, despite operating in physically exposed deployment settings that motivate RBLWE's use. On the software side, Buchmann et al. established the first lightweight RBLWE implementation on 8-bit AVR and Cortex-M0 targets~\cite{buchmann2016high}, while Ebrahimi and Bayat-Sarmadi extended this line with fault-resilient binary Ring-LWE for constrained devices~\cite{ebrahimi2020lightweight}. Both target unaccelerated, general-purpose cores: neither exploits 32-bit SIMD-style instructions nor on-die security peripherals now standard on commodity Cortex-M MCUs. Power-based leakage assessment and binary-level verification of countermeasure survival were also absent from those works. The result is a gap between two otherwise mature lines of work: (i) architecture-aware acceleration is evaluated without implementation-security scrutiny, and (ii) implementation-hardened software is evaluated without architectural acceleration.

This paper closes that gap on a single commodity Cortex-M33 MCU. We map RBLWE's binary polynomial arithmetic onto the core's byte-lane SIMD instructions, packing four coefficients per 32-bit word and using \texttt{UADD8}/\texttt{USUB8} to process them concurrently, which drives most of the encryption and decryption speedup, and pair this with word-oriented encoding and decoding to reduce scalar loads and stores. We also draw randomness from the MCU's Secure Element True Random Number Generator (TRNG) in larger batches rather than one small call at a time, which similarly dominates the key-generation gain by reducing its invocation frequency. Starting from this accelerated Chosen-Plaintext-Attack (CPA) secure core, we add protections in stages: (i) constant-time execution, (ii) hardening against zeroing, random-corruption, and instruction-skip faults, (iii) a Fujisaki--Okamoto (FO)-style adaptive Chosen-Ciphertext-Attack (CCA2) transform built on Secure-Element SHA-256, and (iv) first-order shared CPA decryption, so that each protection layer's cost is individually visible. We further verify, at the linked-binary level, that the compiler preserves these properties, and in doing so identify a \textbf{compiler-induced masking flaw} that motivates a Thumb-assembly replacement. In summary, this paper makes the following contributions (also summarized in Fig.~\ref{fig:main_block_diagram}):

\begin{figure}[t]
\centering
\includegraphics[width=\linewidth]{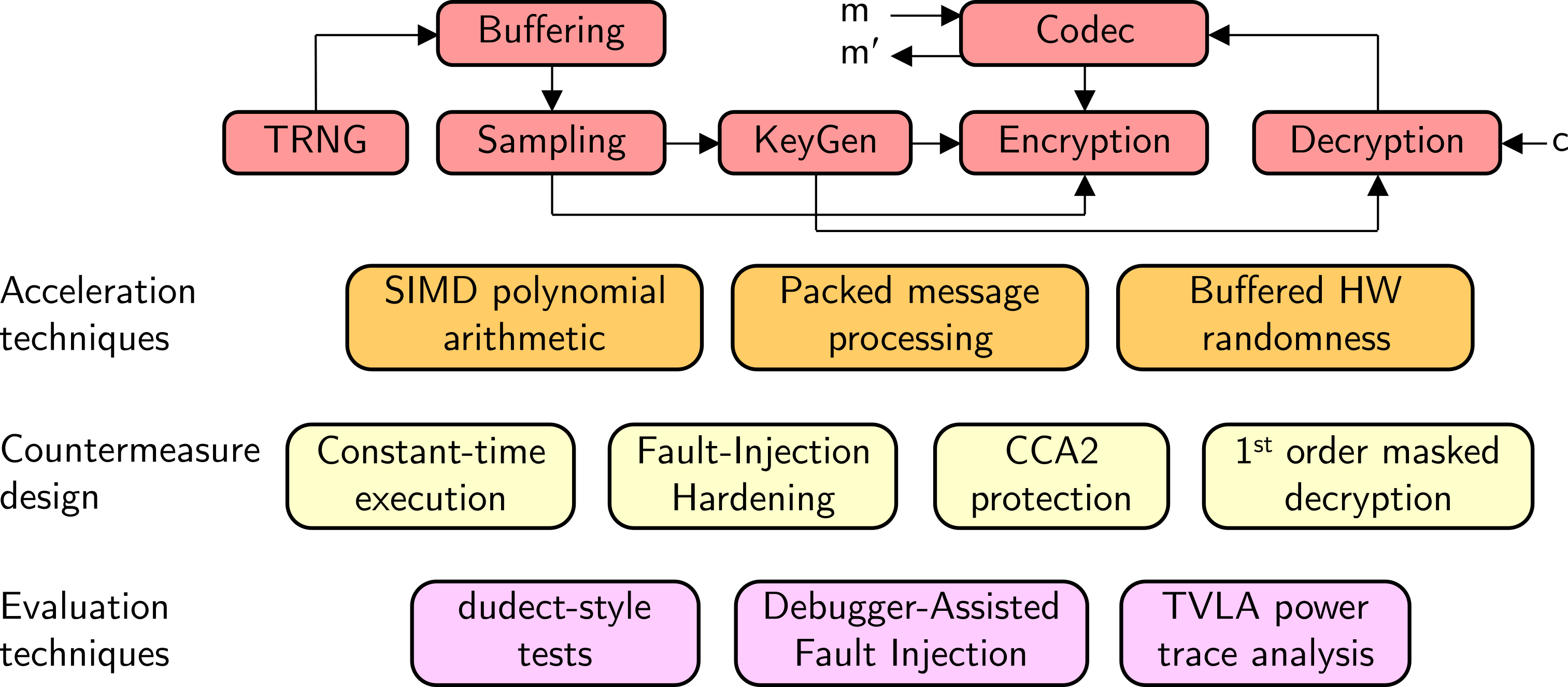}
\caption{End-to-end overview of the proposed RBLWE pipeline (top), the three acceleration techniques and four staged countermeasures applied to it, and the evaluation techniques used to validate them.}
\label{fig:main_block_diagram}
\end{figure}
\begin{enumerate}[label=\textbf{C\arabic*}., leftmargin=*, itemsep=2pt, topsep=2pt]
    \item \textbf{ISA- and Hardware-Assisted RBLWE Acceleration:} Byte-lane SIMD arithmetic, packed encoding/decoding, and a buffered Secure-Engine TRNG together give same-core speedups of $4.06\times$, $3.35\times$, and $3.01\times$ for key generation, encryption, and decryption over a scalar baseline, and $3.52\times$/$3.18\times$ lower encryption/decryption cycle counts than the reference Cortex-M0 implementation; SIMD arithmetic accounts for most of the encryption and decryption gain, while buffered TRNG batching accounts for most of the key-generation gain.
    \item \textbf{Staged Implementation Hardening:} The accelerated core is progressively extended with constant-time execution, fault-hardened comparison and selection, a CCA2-oriented wrapper, and first-order shared CPA decryption, with the cycle and memory cost of each layer reported separately.
    \item \textbf{Binary-Level Verification:} Final Executable and Linkable Format (ELF) inspection confirms that countermeasures survive compilation, and identifies a case where GCC recombined masked shares inside a \texttt{noinline} gadget, resolved with a hand-written assembly replacement.
    \item \textbf{Implementation-Security Evaluation:} Dudect-style timing tests, component-level Test Vector Leakage Assessment (TVLA), and 2,678 debugger-assisted fault trials provide implementation-level evidence for the staged protections, including cases where an injected fault propagated through despite the tested countermeasures.
\end{enumerate}

The rest of the paper is organized as follows. Section~\ref{sec:background} presents the RBLWE background while Section~\ref{sec:optimized_baseline} details the proposed architecture-aware acceleration. Section~\ref{Sec3} defines the adversary model and then presents the staged countermeasures. Section~\ref{sec:methodology} describes the evaluation methodology, and the results are discussed in Section~\ref{sec:results}. Section~\ref{sec:conclusion} concludes this article.

\section{Background and Preliminaries}
\label{sec:background}
This section summarizes the Ring-LWE foundations underlying the RBLWE construction and defines the binary distributions, encryption/decryption operations, and bias-aware decoding rule used in this paper.

\subsection{Ring-LWE Computational Model}
Learning With Errors (LWE) is a well-established hardness assumption: recovering a secret from noisy linear relations is, in the average case, at least as hard as solving certain lattice problems in the worst case~\cite{regev2009lattices}. Ring-LWE transfers this structure to a polynomial quotient ring~\cite{lyubashevsky2013ideal}, operating over
\begin{equation}
R_q = \mathbb{Z}_q[x]/(x^n+1),
\end{equation}
where $n$ is the polynomial degree and $q$ is the coefficient modulus. Each ring element is a polynomial of degree at most $n-1$, with coefficients indexed $i=0,\dots,n-1$ and reduced modulo $q$. Addition is coefficient-wise; multiplication is negacyclic, since $x^n=-1$ causes terms crossing degree $n-1$ to wrap to lower-degree positions with a sign change. RBLWE further restricts the secret and error polynomials to binary coefficients, and this coefficient layout, together with the ring's modular and negacyclic structure, determines the polynomial kernels developed in Section~\ref{sec:optimized_baseline}.

\subsection{RBLWE Dataflow and Bias-Aware Decoding}
We adopt the RBLWE public-key encryption scheme of Buchmann et al. \cite{buchmann2016high}, including its coefficient-dependent decoding rule. Binary sampling reduces sampler and arithmetic complexity relative to wider error distributions, but also introduces a coefficient-dependent noise bias that the decoder (Section~\ref{sec:bias-aware-decoding}) must correct for.

\subsubsection{Key generation}
A public polynomial $a$ is selected from $R_q$. Key generation then draws two binary polynomials independently and uniformly, i.e., with each coefficient distributed as $\operatorname{Bernoulli}(1/2)$,
\begin{equation}
r_1,r_2\leftarrow\{0,1\}^{n}
\end{equation}
and computes
\begin{equation}
p = r_1-a r_2 \quad \text{in } R_q.
\end{equation}
The resulting public key is $pk=(a,p)$, while the secret key is $sk=r_2$. As in~\cite{buchmann2016high}, $a$ is treated as a system parameter fixed at deployment rather than transmitted with each public key; our implementation stores a local copy of $a$ inside the in-memory public-key object rather than as a separate shared value.

\subsubsection{Encryption}
Given a public key $pk=(a,p)$ and a binary message polynomial $m$, encryption draws three fresh binary polynomials independently and uniformly from $\{0,1\}^{n}$,
\begin{equation}
e_1,e_2,e_3\leftarrow\{0,1\}^{n}.
\end{equation}
It then computes the two ciphertext components
\begin{align}
c_1 &= a e_1 + e_2, \\
c_2 &= p e_1 + e_3 + \operatorname{Encode}(m),
\end{align}
where all operations are performed in $R_q$ and $q$ is even. Each message bit is mapped to the corresponding coefficient of the encoded polynomial:
\begin{equation}
\operatorname{Encode}(m)_i=
\begin{cases}
0, & m_i=0,\\
q/2, & m_i=1.
\end{cases}
\end{equation}
The resulting ciphertext is $c=(c_1,c_2)$.

\subsubsection{Decryption and residual noise}
Given $c=(c_1,c_2)$ and the secret key $r_2$, decryption first computes
\begin{equation}
w = c_1r_2+c_2.
\end{equation}
Substitution yields,
\begin{align}
w &= \operatorname{Encode}(m)+\nu, &
\nu &= e_1r_1+e_2r_2+e_3,
\end{align}
where the $ae_1r_2$ terms cancel and $\nu$ is the residual noise.

\subsubsection{Bias-aware decoding}
\label{sec:bias-aware-decoding}

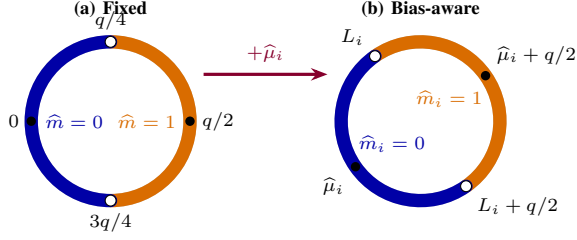
\begin{figure}[t]
\centering
\begin{tikzpicture}[font=\scriptsize]
  \def\R{1.05}
  \tikzset{
    ringzero/.style={blue!65!black, line width=5pt, line cap=round},
    ringone/.style={orange!85!black, line width=5pt, line cap=round},
    ringpoint/.style={circle, fill=black, inner sep=1.25pt},
    ringboundary/.style={circle, draw=black, fill=white, inner sep=1.45pt}
  }

  \begin{scope}[xshift=-2.05cm]
    \node[font=\scriptsize\bfseries] at (0,1.48) {(a) Fixed};
    \draw[gray!45, line width=5pt] (0,0) circle (\R);
    \draw[ringone] (-90:\R) arc[start angle=-90,end angle=90,radius=\R];
    \draw[ringzero] (90:\R) arc[start angle=90,end angle=270,radius=\R];
    \node[ringpoint] at (180:\R) {};
    \node[ringpoint] at (0:\R) {};
    \node[ringboundary] at (90:\R) {};
    \node[ringboundary] at (270:\R) {};
    \node[left=1pt] at (180:\R) {$0$};
    \node[right=1pt] at (0:\R) {$q/2$};
    \node[above=1pt] at (90:\R) {$q/4$};
    \node[below=1pt] at (270:\R) {$3q/4$};
    \node[blue!65!black] at (-0.48,0) {$\widehat m=0$};
    \node[orange!85!black] at (0.48,0) {$\widehat m=1$};
  \end{scope}

  \draw[-{Stealth[length=2.2mm]}, very thick, purple!70!black]
    (-0.82,0.62) -- node[above=1pt,font=\scriptsize] {$+\widehat{\mu}_i$} (0.82,0.62);

  \begin{scope}[xshift=2.05cm]
    \node[font=\scriptsize\bfseries] at (0,1.48) {(b) Bias-aware};
    \draw[gray!45, line width=5pt] (0,0) circle (\R);
    \draw[ringone] (-55:\R) arc[start angle=-55,end angle=125,radius=\R];
    \draw[ringzero] (125:\R) arc[start angle=125,end angle=305,radius=\R];

    \node[ringpoint] at (215:\R) {};
    \node[ringpoint] at (35:\R) {};
    \node[ringboundary] at (125:\R) {};
    \node[ringboundary] at (305:\R) {};
    \node[below left=1pt] at (215:\R) {$\widehat{\mu}_i$};
    \node[above right=1pt] at (35:\R) {$\widehat{\mu}_i+q/2$};
    \node[above left=1pt] at (125:\R) {$L_i$};
    \node[below right=1pt] at (305:\R) {$L_i+q/2$};
    \node[blue!65!black] at (-0.38,-0.30) {$\widehat m_i=0$};
    \node[orange!85!black] at (0.38,0.30) {$\widehat m_i=1$};
  \end{scope}
\end{tikzpicture}
\caption{Fixed and bias-aware decoding regions~\cite{buchmann2016high}. The
rotation uses $\widehat{\mu}_i=i-n/2+2$ and
$L_i=\widehat{\mu}_i+q/4$.}
\label{fig:bias_aware_decode_circle}
\end{figure}

The binary variables take values in $\{0,1\}$ and therefore have a built-in positive skew rather than the zero-centered noise of Gaussian-based schemes. In addition, negacyclic multiplication introduces positive and negative wrapped terms whose numbers depend on the coefficient index $i$. Consequently, following the analysis of \cite{buchmann2016high}, the expected residual noise is coefficient-dependent:
\begin{equation*}
\mathbb{E}([e_j r_j]_i)=\frac{-n+2i+2}{4},
\quad
\mu_i=\mathbb{E}([\nu]_i)=i-\frac{n}{2}+\frac{3}{2}.
\end{equation*}
For even $n$, the decoder rounds the half-integer center and computes
\begin{align}
\widehat{\mu}_i &= \lceil\mu_i\rceil=i-n/2+2, &
L_i &= \widehat{\mu}_i+q/4, \notag\\
z_i &= (w_i-L_i)\bmod q, &
\hat{m}_i &= 1\oplus\operatorname{MSB}_{\log_2 q}(z_i).
\label{eq:rblwe_decode}
\end{align}
Thus, subtracting $L_i$ compensates for index-dependent noise bias before the branch-free Most Significant Bit (MSB) decision is applied as visualized in Fig.~\ref{fig:bias_aware_decode_circle}.

\section{Architecture-Aware Implementation}
\label{sec:optimized_baseline}

\subsection{Target Resources and Data Representation}
The implementation runs on the 32-bit ARM Cortex-M33 core of a Silicon Labs EFM32PG28. Polynomial arithmetic remains software-executed, since the on-die Secure Element supplies random bytes but does not accelerate polynomial multiplication. For the evaluated $(n,q)$ = (256, 256) baseline, the coefficients are stored contiguously as bytes. This representation permits four coefficients to occupy one 32-bit register, and the byte-lane arithmetic detailed in Section~\ref{sec:acceleration_mechanisms} keeps each lane's overflow independent, making per-lane wraparound equivalent to reduction modulo 256.

The firmware exposes $(n,q)$ = (256, 256), (256, 128), and (512, 256) as compile-time configurations, matching the parameter sets for which Buchmann et al. reported classical security estimates and per-bit decryption-failure probabilities of 84 bits$/2^{-32}$, 94 bits$/2^{-10}$, and 190 bits$/2^{-18}$, respectively~\cite{buchmann2016high}. Set~2 is retained only for baseline consistency and overhead comparison; its reported decryption-failure-rate (DFR) precludes a deployment recommendation or the CCA2-security claim developed in Section~\ref{sec:countermeasure-design}.

\subsection{Acceleration Mechanisms}
\label{sec:acceleration_mechanisms}

\begin{figure}[!t]
\centering
\includegraphics[width=\linewidth]{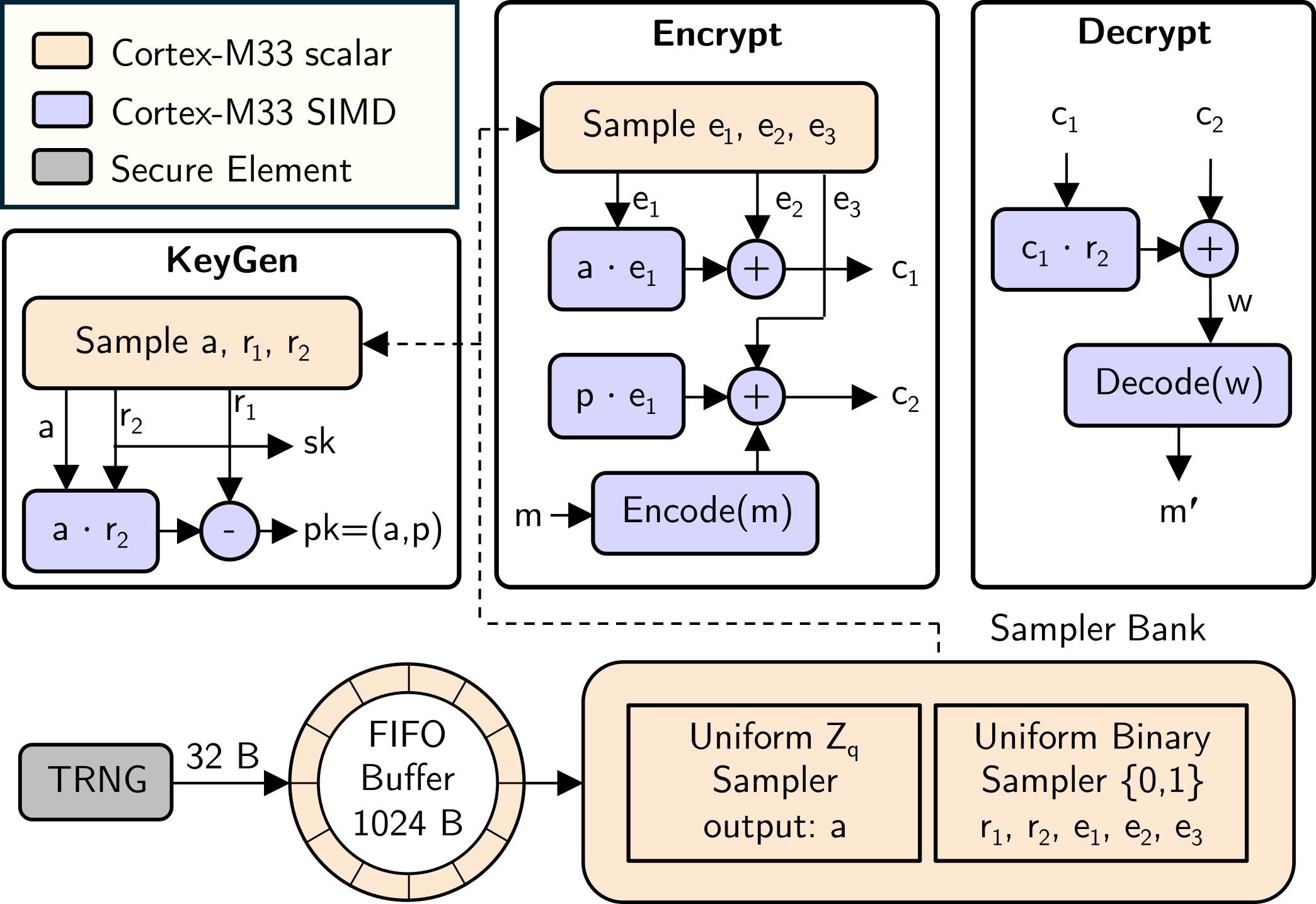}
\caption{Baseline CPA RBLWE dataflow on the EFM32PG28, resolved down to KeyGen, Encrypt, Decrypt, and the shared TRNG-backed Sampler Bank. Packed message encoding and decoding run on the same four-lane SIMD engine as the polynomial arithmetic.}
\label{fig:accelerated_baseline_architecture}
\end{figure}

In our proposed approach, three changes produce the accelerated baseline:
\begin{enumerate}[leftmargin=*,label=\arabic*)]
   \item \textbf{Four-lane polynomial arithmetic:} Algorithm~\ref{alg:simd-negacyclic} maps binary negacyclic multiplication to the Cortex-M33 byte-lane DSP instructions. For each $b_i=1$ and output index $k\in\{0,\ldots,n-1\}$, $r_k$ receives $+a_{k-i}$ when $k\ge i$ and $-a_{n+k-i}$ when $k<i$, as required by $x^n=-1$. Equal-length ranges are processed four coefficients at a time; only their zero-to-three coefficient tails are scalar. Byte-lane wraparound gives reduction modulo 256, followed by a final seven-bit mask when $q=128$.
   \begin{algorithm}[t]
    \footnotesize
\caption{SIMD-Accelerated Negacyclic Multiplication}
\label{alg:simd-negacyclic}
\begin{algorithmic}[1]
\Require $a\in\mathbb Z_q^n$, $b\in\{0,1\}^{n}$, $q\in\{128,256\}$; 8-bit coefficients
\Ensure $r=a\circledast b\in R_q$
\State $r\gets0$
\For{$i\gets0$ to $n-1$}
    \If{$b_i=1$}
        \State \Call{AddRange4}{$r_{i:n-1},a_{0:n-i-1}$}
        \Comment{\texttt{UADD8} + scalar tail}
        \State \Call{SubRange4}{$r_{0:i-1},a_{n-i:n-1}$}
        \Comment{\texttt{USUB8} + scalar tail}
    \EndIf
\EndFor
\If{$q=128$} \State $r\gets r\mathbin{\&}\texttt{0x7f}$ coefficient-wise \EndIf
\State \Return $r$
\end{algorithmic}
\end{algorithm}
    \item \textbf{Packed message processing:}
      the codec processes one message byte as eight coefficients (Algorithm~\ref{alg:packed-codec}). Encoding maps each bit branch-free to $0$ or $q/2$ and stores the results as two 32-bit words; decoding applies (\ref{eq:rblwe_decode}) to eight coefficients and packs the recovered bits into one byte. This preserves the scalar codec output while reducing control and memory overhead; unlike polynomial arithmetic, it is a data-layout optimization and does not use \texttt{\_\_UADD8}/\texttt{\_\_USUB8}. Here, \Call{PackStore}{} is our notation for concatenating eight bytes into two 32-bit words, not an ARM/CMSIS intrinsic.
    \begin{algorithm}[t]
    \footnotesize
    \caption{Packed message encoding and decoding}
    \label{alg:packed-codec}
    \begin{algorithmic}[1]
    \Procedure{PackedEncode}{$m,n,q$}
        \For{$j\gets0$ to $n/8-1$}
            \For{$t\gets0$ to $7$}
                \State $\mu_t\gets\bigl(0-((m_j\gg t)\mathbin{\&}1)\bigr)
                       \mathbin{\&}(q/2)$
            \EndFor
            \State $P_{8j:8j+7}\gets
                   \Call{PackStore}{\mu_0,\ldots,\mu_7}$
        \EndFor
        \State \Return $P$
    \EndProcedure
    \Statex
    \Procedure{PackedDecode}{$P,n,q$}
        \State $\widehat m\gets n/8$ zero bytes
        \For{$i\gets0$ to $n-1$}
            \State $L_i\gets i-n/2+q/4+2$
            \State $z\gets(P_i-L_i)\mathbin{\&}(q-1)$
            \State $d\gets1\oplus\bigl(z\gg(\log_2q-1)\bigr)$
            \State $\widehat m_{\lfloor i/8\rfloor}\gets
                   \widehat m_{\lfloor i/8\rfloor}
                   \mathbin{|}(d\ll(i\bmod8))$
        \EndFor
        \State \Return $\widehat m$
    \EndProcedure
    \end{algorithmic}
    \end{algorithm}
    \item \textbf{Buffered hardware randomness:} a circular pool batches Secure Element TRNG output and supplies the binary samplers. In the prefetched mode, the pool is filled before the timed operation and the pool-refill step is moved off the latency-critical path. In the reported cold start mode, the operation begins with an empty pool, and every required refill is counted against the reported cycle count.
\end{enumerate}
These transformations change the execution mapping, not the RBLWE equations or Bernoulli distributions. The $q=128$ configuration additionally masks the results with \texttt{0x7f} to restrict the coefficient to the valid 7-bit range $[0,127]$, since byte-lane overflow naturally wraps modulo $256$, rather than modulo $128$. This final reduction is valid because $128$ divides $256$. Parameter-specific correctness and the security results are therefore reported separately in Section~\ref{sec:results}. Fig.~\ref{fig:accelerated_baseline_architecture} consolidates the verified baseline dataflow and its mapping to the target resources.

\section{Security Countermeasures}
\label{Sec3}
This section first defines the adversary model assumed for a commodity IoT deployment, then presents the security countermeasures introduced in response.

\subsection{Adversary Model}
\label{sec:threat_model}

\subsubsection{Adversary Capabilities}

In our adopted model, the adversary may request encryptions of chosen plaintexts and submit arbitrary, including malformed, ciphertexts for decryption, observing outputs and externally visible execution behavior but not the secret key or internal random state. Given physical access, the adversary additionally collects timing and power traces across repeated executions and injects one targeted fault per execution, namely instruction skip, state zeroing, or value substitution, at key generation, encryption, decryption, re-encryption, or verification/output stages. A fault is security-relevant if it admits a weak key or malformed ciphertext, releases unauthenticated or corrupted plaintext, or silently corrupts a valid output; explicit failure and implicit rejection remain fail-closed. Underlying RBLWE, hash, and key-derivation primitives are assumed computationally secure.

\subsubsection{Validation Scope}

Evaluation is limited to the tested binaries and components: fixed-versus-random timing, TVLA, and debugger-assisted single faults. Multi-target faults and electromagnetic or invasive injection are considered out of scope for this work.


\subsection{Countermeasure Design}
\label{sec:countermeasure-design}

Building on fault-resilient Binary Ring-LWE~\cite{ebrahimi2020lightweight}, we
add constant-time execution, fault hardening, a CCA2-oriented transform, and
first-order shared CPA decryption as security countermeasures to be tested in our work. 

\subsubsection{Constant-Time Execution}

The baseline binary polynomial multiplication can skip an outer-loop iteration when a sensitive multiplier coefficient is zero. Such skip-zero behavior makes execution time and instruction activity depend on the Hamming weight and positions of the binary multiplier. Let $b_i=b[i]\in\{0,1\}$ denote its $i$th coefficient; depending on the call site, $b$ is a secret-key or ephemeral polynomial. In the protected implementation, the skip branch is replaced by a full-width mask
\begin{equation}
M_i = -(b_i \mathbin{\&} 1),
\end{equation}
so that $M_i=\texttt{0x00000000}$ when $b_i=0$ and $M_i=\texttt{0xFFFFFFFF}$ when $b_i=1$.
The firmware does not materialize the complete negacyclic shift $\operatorname{NShift}_i(a)$. Instead, during outer iteration $i$, it applies $M_i$ to each packed four-coefficient source word loaded from $a$. Nonwrapped words are accumulated with \texttt{\_\_UADD8}, whereas wrapped words are accumulated with \texttt{\_\_USUB8} to implement $x^n\equiv-1$; residual boundary coefficients use the corresponding 8-bit mask. Thus, each masked word is a packed source block contributing to $\operatorname{NShift}_i(a)$, not the complete shifted polynomial. Every $b_i$ follows the same loop structure: zero is accumulated when $b_i=0$, and the $i$th negacyclic shift of $a$ is accumulated when $b_i=1$.

\subsubsection{Fault-Injection Hardening}
If any step of key generation fails, for example a sampling or Secure Element error, both the public and secret key objects are erased before the call returns, rather than leaving a partially formed or possibly weak key pair in memory. During decryption, forward and reverse comparisons produce independent masks, and candidate data is released only if both masks pass and no internal error occurs (Algorithm~\ref{alg:redundant-ct-verify}). Unlike the comparison step, the final output loop and store are not similarly duplicated, leaving a single, non-redundant point that Section~\ref{sec:results} shows can still be exploited.

\begin{algorithm}[t]
\footnotesize
\caption{Redundant verification and hardened output selection}
\label{alg:redundant-ct-verify}
\begin{algorithmic}[1]
\Require received data $C$, recomputed data $\widehat{C}$, candidate $m$,
alternative output $r$, operation status $s$
\State $d_f \gets \mathsf{CompareForward}(C,\widehat{C})$
\State $d_r \gets \mathsf{CompareReverse}(C,\widehat{C})$
\State $M_f \gets \mathsf{EqMask}(d_f)$
\State $M_r \gets \mathsf{EqMask}(d_r)$
\If{$s\neq0$}
    \State $M_f\gets0$; $M_r\gets0$
\EndIf
\For{$i\gets0$ to $|m|-1$}
    \State $out_i\gets r_i\oplus
    ((m_i\oplus r_i)\mathbin{\&}M_f\mathbin{\&}M_r)$
\EndFor
\State \Return $out$
\end{algorithmic}
\end{algorithm}

\subsubsection{CCA2-Oriented Protection}

\begin{figure}[t]
\centering
\definecolor{ccaNavy}{HTML}{17324D}
\definecolor{ccaBlue}{HTML}{2F6B9A}
\definecolor{ccaTeal}{HTML}{16877B}
\definecolor{ccaGreen}{HTML}{4B8F55}
\definecolor{ccaAmber}{HTML}{C98216}
\definecolor{ccaRed}{HTML}{B6544C}
\definecolor{ccaSlate}{HTML}{536273}
\definecolor{ccaLine}{HTML}{B8C3CD}
\definecolor{ccaPaper}{HTML}{F5F7F9}
\tikzset{
  ccaEncHead/.style={fill=ccaBlue, rounded corners=2pt,
    minimum width=42mm, minimum height=5.2mm, inner sep=1pt,
    align=center, font=\sffamily\footnotesize\bfseries, text=white},
  ccaDecHead/.style={fill=ccaTeal, rounded corners=2pt,
    minimum width=42mm, minimum height=5.2mm, inner sep=1pt,
    align=center, font=\sffamily\footnotesize\bfseries, text=white},
  ccaCard/.style={draw=ccaLine, fill=white, rounded corners=2.2pt,
    text width=38mm, minimum width=42mm, inner xsep=2mm,
    inner ysep=1.5mm, align=left, font=\sffamily\scriptsize,
    text=ccaNavy, line width=0.55pt},
  ccaInput/.style={ccaCard, fill=ccaPaper, draw=ccaSlate!55},
  ccaHash/.style={ccaCard, fill=ccaTeal!5, draw=ccaTeal!65},
  ccaCompute/.style={ccaCard, fill=ccaBlue!5, draw=ccaBlue!65},
  ccaVerify/.style={ccaCard, fill=ccaAmber!7, draw=ccaAmber!75},
  ccaArrow/.style={-{Stealth[length=1.5mm,width=1.05mm]},
    draw=ccaSlate, line width=0.55pt, rounded corners=1pt}
}
\begin{tikzpicture}[node distance=1.35mm and 3mm]
  \node[ccaEncHead] (eh) {(a) Encryption};
  \node[ccaDecHead, right=2mm of eh] (dh) {(b) Decryption};

  \node[ccaInput, below=1.25mm of eh] (ein)
    {\textbf{\textcolor{ccaSlate}{1\enspace Inputs}}\\[-0.3mm]
     $m$; fresh $v$ from software TRNG pool\\
     $hpk=H(pk)$ from SE SHA-256};
  \node[ccaHash, below=of ein] (ederive)
    {\textbf{\textcolor{ccaTeal}{2\enspace Derive coins and mask}}\\[-0.3mm]
     SE SHA-256 with software framing\\
     $G(v\parallel hpk)$};
  \node[ccaCompute, below=of ederive] (ecore)
    {\textbf{\textcolor{ccaBlue}{3\enspace RBLWE and message path}}\\[-0.3mm]
     $c_1=ae_1+e_2$\\
     $c_2=pe_1+e_3+\operatorname{Encode}(v)$\\
     $c_3=m\oplus G(v\parallel hpk)$};
  \node[ccaHash, below=of ecore] (etag)
    {\textbf{\textcolor{ccaTeal}{4\enspace Authenticate and transmit}}\\[-0.3mm]
     SE SHA-256 context tag $c_4$\\
     $C=(c_1,c_2,c_3,c_4)$};

  \node[ccaInput, below=1.25mm of dh] (din)
    {\textbf{\textcolor{ccaSlate}{1\enspace Inputs}}\\[-0.3mm]
     $C$, shared $sk'$, and $hpk$\\
     $z$ from software TRNG pool at key setup};
  \node[ccaCompute, below=of din] (drecover)
    {\textbf{\textcolor{ccaBlue}{2\enspace Recover vector}}\\[-0.3mm]
     Cortex-M33 shared CPA decryption\\
     $v'=\operatorname{Dec}_{sk'}(c_1,c_2)$};
  \node[ccaHash, below=of drecover] (dregen)
    {\textbf{\textcolor{ccaTeal}{3\enspace Recover candidate}}\\[-0.3mm]
     SE: $G(v'\parallel hpk)$\\
     M33: $m'=c_3\oplus G(v'\parallel hpk)$};
  \node[ccaVerify, below=of dregen] (dcheck)
    {\textbf{\textcolor{ccaAmber}{4\enspace Recompute ciphertext}}\\[-0.3mm]
     SE $H,HH$: coins; M33: $c'_1,c'_2$\\
     SE: $c'_4$ and $R(z,H(C))$};
  \node[ccaVerify, below=of dcheck] (dout)
    {\textbf{\textcolor{ccaAmber}{5\enspace Verify and select}}\\[-0.3mm]
     Bidirectional compare and two-mask Thumb assembly selection\\[0.25mm]
     \textcolor{ccaGreen}{\textbf{valid}} $\rightarrow m'$\\
     \textcolor{ccaRed}{\textbf{invalid}} $\rightarrow R(z,H(C))$\\[-0.2mm]
     \textcolor{ccaRed}{\textbf{internal failure}} $\rightarrow$ error};

  \draw[ccaArrow] (ein) -- (ederive);
  \draw[ccaArrow] (ederive) -- (ecore);
  \draw[ccaArrow] (ecore) -- (etag);
  \draw[ccaArrow] (din) -- (drecover);
  \draw[ccaArrow] (drecover) -- (dregen);
  \draw[ccaArrow] (dregen) -- (dcheck);
  \draw[ccaArrow] (dcheck) -- (dout);
\end{tikzpicture}
\caption{Implemented FO-style CCA2 RBLWE dataflow on the EFM32PG28. Blue, teal, amber, and gray cards denote Cortex-M33 computation, Secure Element hashing, verification and selection, and inputs, respectively. The TRNG and SHA-256 engine are hardware services. Authentication failure returns $R(z,H(C))$; an internal service failure reports an error.}
\label{fig:cca2_rblwe_flow}
\end{figure}
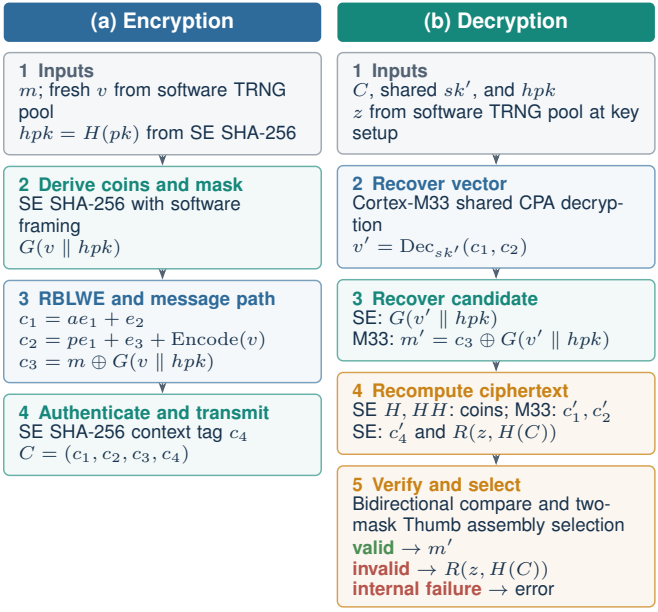

The CCA2-oriented layer is designed to prevent a modified ciphertext from being accepted as valid. It follows the deterministic re-encryption approach used by Ebrahimi and Bayat-Sarmadi~\cite{ebrahimi2020lightweight} and Oder et al.~\cite{oder2018practical}. As shown in Fig.~\ref{fig:cca2_rblwe_flow}, RBLWE encrypts a fresh binary vector $v$ rather than the message $m$. The message is masked in $c_3$, while $c_4$ authenticates the complete ciphertext. Here, $H$ denotes SHA-256 hashing, including the domain-separated seed derivation from $(v,m,hpk)$; $HH$ is a domain-separated SHA-256 chaining function for the subsequent seeds; and $G$ is an HMAC-based Key Derivation Function (HKDF) expansion using SHA-256 that produces the plaintext mask. For Set~1 ($n=256$, $q=256$), the four components of $C=(c_1,c_2,c_3,c_4)$ occupy 256, 256, 32, and 32 bytes, respectively; $c_3$ and $c_4$ scale with $n$, giving 64 bytes each at Set~3 ($n=512$).

Decryption first recovers candidates $v'$ and $m'$, then rebuilds $c'_1,c'_2,c'_4$. The candidate message is returned only when all three rebuilt components match the received ones. A mismatch returns the pseudorandom message $R(z,H(C))$  instead and does not expose a validity flag. Furthermore, internal processing failures, such as a failed Secure Element hash call, are reported as errors.

Firmware derives several independent functions, $H$, $G$, and the tag/coin derivations, from the Secure Element's single SHA-256 hardware. Each call is tagged with a distinct label and its output combined via HKDF, keeping the functions cryptographically separate. The CCA2 wrapper follows the same re-encryption pattern as standardized schemes like ML-KEM but uses its own byte layout, so the two are not interchangeable. Like the underlying RBLWE scheme, the wrapper is evaluated empirically here, not backed by a formal CCA2 proof.

\subsubsection{First-Order Shared CPA Decryption}

The countermeasures introduced so far leave the core polynomial arithmetic exposed to power analysis: an attacker could still recover $r_2$ (secret key) from repeated observations of the ciphertext-times-key multiplication. Masking closes this gap by splitting every sensitive value into two random shares, so no single observed value depends on the whole secret. Masking begins when $r_2$ is created, not at the decoder. Each binary coefficient of $r_2$ is stored as two arithmetic shares,
\begin{equation}
r_2=r_{2,0}+r_{2,1}\pmod q.
\end{equation}
The first share is random and the second completes the required binary value. Before every decryption, a fresh random polynomial $\delta$ refreshes the key as $r_{2,0}\leftarrow r_{2,0}+\delta$ and $r_{2,1}\leftarrow r_{2,1}-\delta$, leaving their sum unchanged.

The firmware then computes $u_0=c_1r_{2,0}$ and $u_1=c_1r_{2,1}$, followed by $w_0=u_0$ and $w_1=u_1+c_2$. Hence,
\begin{equation}
w_0+w_1=c_1r_2+c_2,
\end{equation}
without first reconstructing either $r_2$ or the decryption polynomial $w$. The shared-key build likewise computes the public-key product from both shares without reconstructing $r_2$. Because the shares are general elements of $\mathbb{Z}_q$, these operations use the general rather than the faster binary multiplier, which dominates the shared KeyGen overhead.

Unlike the four-lane byte-SIMD arithmetic used elsewhere, the decoder processes each coefficient pair $(w_{i,0},w_{i,1})$ separately, since recovering a single bit from a masked value requires a carry-propagating conversion rather than independent, byte-lane-parallel operations. This conversion turns the arithmetic shares into two Boolean shares via a masked $\log_2 q$-bit ripple-carry addition, protecting each carry step with an Ishai-Sahai-Wagner (ISW)-style masked-AND gadget that computes on shares alone, then applies the public decoding offset without combining the shares. For Set~1, with $(n,q)=(256,256)$ and uniform binary sampling, $L_i=i-62$ and the decoded bit is
\begin{equation}
m_i=1\oplus\operatorname{MSB}((w_i-L_i)\bmod256).
\end{equation}
The output bit shares are packed into two $n/8$-byte buffers and combined only after all $n$ coefficients have been decoded.

Each coefficient consumes five buffered random bytes: two input masks, two bytes for arithmetic-to-Boolean conversion, and one byte for the public-offset addition. The decoder and key-share refresh therefore consume $6n$ random bytes per decryption (1,536 bytes for Set~1). For Set~1, the source-level path also holds four 256-byte polynomial temporaries, a 256-byte refresh buffer, and two 32-byte message-share buffers. A random-pool failure clears the output and returns an error. Masking ends when the two message buffers are combined; the subsequent CCA2 hashing, re-encryption, comparison, and output selection remain unmasked. The implemented claim is therefore first-order shared CPA decryption, not a fully masked CCA2-PKE operation.

\section{Experimental Setup and Methodology}
\label{sec:methodology}

\begin{figure}[t]
\centering
\includegraphics[width=\columnwidth]{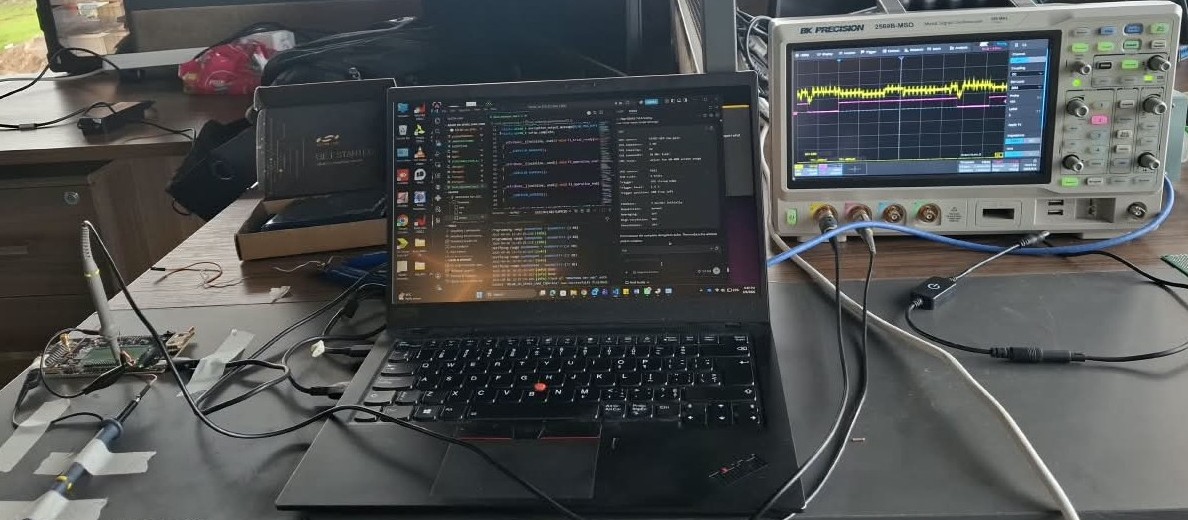}
\caption{TVLA acquisition setup: the EFM32PG28 target board, host-controlled capture software, and BK Precision 2569B-MSO oscilloscope recording the AEM current signal and GPIO trigger.}
\label{fig:tvla-experimental-setup}
\end{figure}

\subsection{Evaluated Configurations}

Each parameter tuple is rebuilt with uniform binary sampling in seven cumulative configurations: scalar arithmetic and codec with direct TRNG calls, packed arithmetic, packed codec, buffered branchy CPA, branch-free CPA, branch-free plus CCA2/fault protection, and the latter with first-order shared CPA decryption. End-to-end measurements begin without prefetched randomness; the sampler and baseline arithmetic remain fixed after the accelerated CPA boundary, and CCA2 processing remains unmasked.

\subsection{Toolchain Constraints and Binary-Level Verification}

Optimizing compilers do not necessarily preserve source-level constant-time or masking properties \cite{schneider2025breaking}. The evaluated firmware uses ARM GNU Toolchain GCC 12.2.1 (build 20221205), targets Thumb code for the Cortex-M33, and is compiled as C17 with \texttt{-Os} and \texttt{-fno-lto}. Compiler identity, options, and source revision are therefore part of the evaluated configuration; changing any of them requires the binary audit to be repeated.

Using \texttt{arm-none-eabi-objdump} and \texttt{arm-none-eabi-nm} from the same toolchain, we inspected the application objects and final ELF file for secret-dependent branches and memory addresses, fixed loop bounds, removal of countermeasure operations, and recombination of Boolean shares. The audit was performed on the fully hardened CCA2-masked configuration for parameter Set 1 (\(n=256,\ q=256\)), with branch-free binary multiplication, CCA2-PKE, and masked decryption enabled. The audit confirmed mask-based execution in the binary multiplier, fixed control flow in the masked decoder, full-length ciphertext comparison, and branch-free implicit-rejection selection. The remaining loop branches depend only on fixed public iteration counts.

\subsection{Performance and Memory Measurement}
The Cortex-M33 Data Watchpoint and Trace (DWT) cycle counter measures key generation, encryption, or decryption with interrupts disabled, and correctness is checked outside the timed interval. Measurements distinguish operation latency when the required entropy has already been prefetched into the TRNG-backed pool from caller-visible end-to-end latency when execution begins with an empty pool and, therefore, includes synchronous hardware refilling. Firmware builds without masked decryption use a fixed 1,024-byte pool. When masked decryption is enabled, the secret polynomial is stored and processed as two arithmetic shares, $s=s_0+s_1 \pmod q$, and the pool capacity is increased to $6n$ bytes so that all randomness required by one protected decryption can be prefetched. Reported cold-start values include batched synchronous TRNG refill, pool reads, polynomial arithmetic, Secure-Element SHA-256 calls, and countermeasure logic, but exclude test-vector preparation, correctness checks, and UART output. This cold-start measurement is deliberately conservative, not a formal worst-case bound: whichever operation runs first has to wait for fresh randomness, and that wait counts against it. Prefetching avoids that wait but does not remove the underlying cost, it only moves the work and energy outside the timed window.

Flash usage is the linked \texttt{.text+data} size. Static RAM is the map file's allocated \texttt{.data} plus ordinary \texttt{.bss}, excluding the linker-reserved stack and memory-manager heap. Reported Stack and Heap figures are those linker allocations, not measured runtime high-water marks.

\subsection{Fixed-versus-Random Timing Tests}

We perform four dudect-style tests covering polynomial multiplication during decryption and encryption, as well as both complete CCA2 operations. Each test collects 100,000 DWT measurements, divided equally between two classes, using the $(n,q)=(256,256)$ firmware image. In the decryption test, a fixed secret binary multiplier is paired with either a fixed or random ciphertext polynomial. In the encryption test, a fixed polynomial is paired with either a fixed or random temporary binary multiplier. Complete encryption compares an all-zero plaintext with random plaintexts under one public key. Complete decryption compares valid ciphertexts with branchlessly corrupted invalid ciphertexts under one secret key. Balanced class labels are shuffled before capture. Both classes use the same untimed preparation, working-buffer addresses, branchless selection, and untimed warm-up. Random generation, pool refill, validation, and UART transmission remain outside the timed call. Offline analysis applies Welch's $t$-test where the no-detection criterion requires final and maximum running $|t|<4.5$ with no sustained growth.

\subsection{Power-Trace Leakage Assessment}

Component-level fixed-versus-random TVLA compares the unshared and first-order shared multiplication kernels for $(n,q)$ = (256, 256). A BK Precision 2569B-MSO oscilloscope records the board's Advanced Energy Monitor (AEM) current-measurement output (Fig.~\ref{fig:tvla-experimental-setup}) on channel~1 at 5~MSa/s; a GPIO trigger pulse on channel~2 delimits the selected kernel. Input preparation, UART synchronization, and random-pool refill occur while the trigger is low. Captures are aligned to the trigger edge and analyzed offline with the $|t|=4.5$ threshold.

\subsection{Debugger-Assisted Fault Campaigns}
\label{sec:fault-campaigns}

On-target logical injections use the GNU Debugger (GDB) through J-Link at addresses resolved from the campaign's linked ELF. Each campaign applies a single intervention per execution: zeroing selected state, substituting seeded values in registers or stack locations, or emulating a Thumb instruction skip by advancing the program counter to the next instruction. Valid or deliberately invalid fixed ciphertexts are selected according to the target. Each run records the API status, output class, candidate-bit release, crash or timeout behavior, and whether the injection point was reached. Zeroing uses three repetitions at each of ten locations. Randomization exhausts all 256 byte values separately for each of the forward and reverse ciphertext-validity masks used by the hardened final plaintext selector, applies 1,000 seeded 32-bit values to each comparison accumulator, and uses 100 seeded values for a recovered intermediate. Instruction-skip experiments use three repetitions at each of 12 security-critical instructions. These experiments represent reproducible, software-controlled fault models and do not constitute claims about any particular voltage, clock, or electromagnetic glitch apparatus.

\section{Results and Discussion}
\label{sec:results}

\subsection{Per-Mechanism Acceleration Gains}

Table~\ref{tab:baseline_acceleration} isolates the three cumulative acceleration changes presented in Section~\ref{sec:acceleration_mechanisms}. SIMD arithmetic provides the main arithmetic gain, delivering $1.51\times$, $2.40\times$, and $2.97\times$ speedups over the scalar build for KeyGen, encryption, and decryption, respectively. Packed encoding and decoding provide smaller incremental gains of $1.006\times$ and $1.025\times$ for encryption and decryption, while leaving KeyGen essentially unchanged. Batched TRNG access reduces Secure Element invocation overhead and adds $2.68\times$ and $1.39\times$ incremental gains to KeyGen and encryption; CPA decryption is unaffected because it requires no randomness. All measurements use an empty TRNG pool, so refills occur synchronously and no gain is attributed to prefetching. Combined, these mechanisms provide $4.06\times$, $3.35\times$, and $3.01\times$ speedups over the scalar Cortex-M33 baseline for KeyGen, encryption, and decryption. Compared with the reference Cortex-M0 implementation, the final Cortex-M33 build achieves $3.52$--$3.75\times$ encryption and $3.18$--$3.62\times$ decryption cycle-count speedups across the three parameter sets.

\begin{table*}[!t]
\centering
\scriptsize
\caption{Baseline acceleration for Set 1 and compact all-set comparison.}
\label{tab:baseline_acceleration}
\begin{adjustbox}{max width=\textwidth}
\begin{tabular}{@{} l l r r r @{}}
\toprule
Configuration & Platform & KeyGen [Kcycles (ms)] & Encrypt [Kcycles (ms)] & Decrypt [Kcycles (ms)] \\
\midrule
Scalar arithmetic + scalar codec + direct SE TRNG
& Cortex-M33, 39 MHz & 783.783 (20.10 ms) & 898.871 (23.05 ms) & 381.322 (9.78 ms) \\
Packed SIMD arithmetic + scalar codec + direct SE TRNG
& Cortex-M33, 39 MHz & 520.450 (13.34 ms) & 373.886 (9.59 ms) & 128.547 (3.30 ms) \\
SIMD + optimized encoding/decoding + direct SE TRNG
& Cortex-M33, 39 MHz & 517.612 (13.27 ms) & 371.505 (9.53 ms) & 125.369 (3.21 ms) \\
SIMD + optimized encoding/decoding + batched TRNG buffer (cold)
& Cortex-M33, 39 MHz & 193.275 (4.96 ms) & 268.056 (6.87 ms) & 126.568 (3.25 ms) \\
\multicolumn{2}{@{}l}{\textbf{Controlled cycle-count speedup, scalar to final}}
& \textbf{4.06$\times$} & \textbf{3.35$\times$} & \textbf{3.01$\times$} \\
\specialrule{1.0pt}{3pt}{3pt}
\multicolumn{5}{@{}l}{\textbf{Accelerated CPA comparison across all tested parameter sets}} \\
\addlinespace[2pt]
Set 1: $(n,q)=(256,256)$ & M33/M0 & 193.275/--
& 268.056/944.000 (\textbf{3.52$\times$})
& 126.568/403.000 (\textbf{3.18$\times$}) \\
Set 2: $(n,q)=(256,128)$ & M33/M0 & 199.583/--
& 266.696/999.000 (\textbf{3.75$\times$})
& 127.523/437.000 (\textbf{3.43$\times$}) \\
Set 3: $(n,q)=(512,256)$ & M33/M0 & 614.738/--
& 979.851/3\,483.000 (\textbf{3.55$\times$})
& 469.875/1\,701.000 (\textbf{3.62$\times$}) \\
\bottomrule
\end{tabular}
\end{adjustbox}
\vspace{2pt}
\begin{minipage}{\textwidth}
\raggedright
\footnotesize
\emph{Note:} Direct and batched TRNG configurations use 4-byte requests and 32-byte refills, respectively. All-set entries report this work/Buchmann et al.~\cite{buchmann2016high} in Kcycles, followed by the cycle-count speedup; ``--'': unreported KeyGen.
\end{minipage}
\end{table*}

\subsection{Binary-Level Verification Finding}
The binary audit in Section~\ref{sec:methodology}, checking for secret-dependent branches, fixed loop bounds, and recombination of Boolean shares, surfaced a compiler-induced flaw in the original C masked-AND gadget. Although marked \texttt{noinline}, GCC factored its expression into an instruction computing $b_0\mathbin{\oplus}b_1$, temporarily reconstructing the sensitive value from its two Boolean shares. We replaced the gadget with standalone Thumb assembly that holds the four ISW cross-products in separate registers. Reinspection confirmed that the linked routine neither recombines the input shares nor introduces conditional branches or secret-dependent memory accesses.

\subsection{Performance Impact of Staged Security Countermeasures}

Table~\ref{tab:rblwe_all_sets_results} reports the impact of branch-free (BF) execution, the combined CCA2/fault layer, and shared CPA decryption for each parameter set. Branch-free execution's uniform, no-skip processing raises the KeyGen cycle count over the branchy baseline even before any further protection is added. CCA2 affects decryption most because it adds recovery, re-encryption, verification, and implicit rejection. In the current shared-key (masked) build, secret sharing leaves encryption essentially unchanged but dominates KeyGen and decryption through general multiplication on arithmetic shares, key refresh, and masked decoding.

Table~\ref{tab:set1_prefetched} gives the companion prefetched measurements
for Set~1. Relative to the cold-start values in
Table~\ref{tab:rblwe_all_sets_results}, the KeyGen reduction falls from 29.6\%
at the branchy CPA baseline to 4.0\% in the final shared configuration, while
final-configuration encryption and decryption improve by 0.9\% and 6.8\%,
respectively. The benefit is therefore operation-dependent and follows the
relative contribution of randomness generation to the timed path.

\begin{table*}[t]
\centering
\scriptsize
\caption{Staged cold-start performance and memory costs across RBLWE parameter sets.}
\label{tab:rblwe_all_sets_results}
\begin{adjustbox}{max width=\textwidth}
\begin{tabular}{@{} l r r r r r r r @{}}
\toprule
Stage & KeyGen [Kcycles (ms)] & Encrypt [Kcycles (ms)] & Decrypt [Kcycles (ms)] & ROM (B) & Static RAM (B) & Stack (B) & Heap (B) \\
\midrule
\multicolumn{8}{@{}l}{\textbf{Set 1: $n=256$, $q=256$}} \\
\addlinespace[2pt]
Branchy CPA baseline
& 193.275 (4.96 ms) & 268.056 (6.87 ms) & 126.568 (3.25 ms) & 26\,612 & 4\,164 & 8\,192 & 249\,788 \\
Branch-free CPA
& 310.189 (7.95 ms) & 498.272 (12.78 ms) & 240.100 (6.16 ms) & 26\,644 & 4\,164 & 8\,192 & 249\,788 \\
Branch-free + CCA2/fault protection
& 330.053 (8.46 ms) & 604.186 (15.49 ms) & 871.055 (22.33 ms) & 29\,048 & 4\,804 & 8\,192 & 249\,148 \\
BF + CCA2/fault + shared CPA decryption
& 2\,661.295 (68.24 ms) & 604.811 (15.51 ms) & 3\,973.488 (101.88 ms) & 30\,200 & 5\,572 & 8\,192 & 248\,380 \\
\midrule
\multicolumn{8}{@{}l}{\textbf{Set 2: $n=256$, $q=128$}} \\
\addlinespace[2pt]
Branchy CPA baseline
& 199.583 (5.12 ms) & 266.696 (6.84 ms) & 127.523 (3.27 ms) & 26\,676 & 4\,164 & 8\,192 & 249\,788 \\
Branch-free CPA
& 313.338 (8.03 ms) & 503.665 (12.91 ms) & 242.374 (6.21 ms) & 26\,676 & 4\,164 & 8\,192 & 249\,788 \\
Branch-free + CCA2/fault protection
& 332.078 (8.51 ms) & 609.302 (15.62 ms) & 879.130 (22.54 ms) & 29\,080 & 4\,804 & 8\,192 & 249\,148 \\
BF + CCA2/fault + shared CPA decryption
& 3\,055.053 (78.33 ms) & 610.505 (15.65 ms) & 4\,315.989 (110.67 ms) & 30\,264 & 5\,572 & 8\,192 & 248\,380 \\
\midrule
\multicolumn{8}{@{}l}{\textbf{Set 3: $n=512$, $q=256$}} \\
\addlinespace[2pt]
Branchy CPA baseline
& 614.738 (15.76 ms) & 979.851 (25.12 ms) & 469.875 (12.05 ms) & 26\,644 & 5\,508 & 8\,192 & 248\,444 \\
Branch-free CPA
& 1\,109.371 (28.45 ms) & 1\,974.466 (50.63 ms) & 969.186 (24.85 ms) & 26\,676 & 5\,508 & 8\,192 & 248\,444 \\
Branch-free + CCA2/fault protection
& 1\,105.646 (28.35 ms) & 2\,065.859 (52.97 ms) & 3\,043.904 (78.05 ms) & 29\,048 & 6\,692 & 8\,192 & 247\,260 \\
BF + CCA2/fault + shared CPA decryption
& 11\,341.880 (290.82 ms) & 2\,065.767 (52.97 ms) & 14\,815.736 (379.89 ms) & 30\,264 & 9\,252 & 8\,192 & 244\,700 \\
\bottomrule
\end{tabular}
\end{adjustbox}
\vspace{2pt}
\begin{minipage}{\textwidth}
\raggedright
\footnotesize
\emph{Note:} Measurements use the 39-MHz Cortex-M33 and report end-to-end cold-start latency, including any required batched synchronous TRNG refill.
\end{minipage}
\end{table*}

\begin{table}[t]
\centering
\caption{Set~1 prefetched latency and reduction from cold start.}
\label{tab:set1_prefetched}
\scriptsize
\setlength{\tabcolsep}{2.5pt}
\begin{adjustbox}{max width=\columnwidth}
\begin{tabular}{@{}lrrr@{}}
\toprule
Stage & KeyGen & Encrypt & Decrypt \\
\midrule
Branchy CPA
& 136.031 (29.6\%) & 243.861 (9.0\%) & -- \\
Branch-free CPA
& 253.568 (18.3\%) & 481.297 (3.4\%) & -- \\
BF + CCA2/fault
& 268.244 (18.7\%) & 598.590 (0.9\%) & -- \\
BF + CCA2/fault + shared
& 2\,554.062 (4.0\%) & 599.188 (0.9\%) & 3\,701.445 (6.8\%) \\
\bottomrule
\end{tabular}
\end{adjustbox}
\vspace{2pt}
\begin{minipage}{\columnwidth}
\footnotesize
\emph{Note:} Kcycles (\% reduction from Table~\ref{tab:rblwe_all_sets_results});
``--'': no randomness to prefetch.
\end{minipage}
\end{table}

\begin{figure}[t]
\centering
\includegraphics[width=0.8\columnwidth]{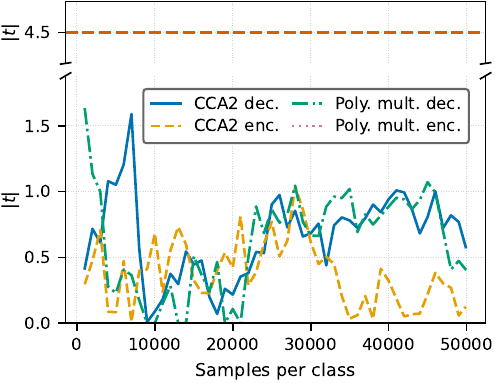}
\caption{Running Welch $|t|$ convergence for the four timing tests; dashed lines mark the $|t|=4.5$ reference.}
\label{fig:dudect-four-tests}
\end{figure}
\vspace{-0.5cm}

\begin{figure}[t]
\centering
\includegraphics[width=0.85\columnwidth]{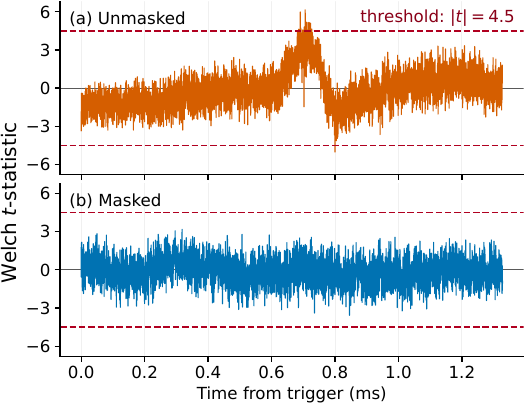}
\caption{Fixed-versus-random TVLA comparison of unmasked (unshared) and first-order masked (shared) multiplication; dashed lines mark $|t|=4.5$.}
\label{fig:masked-unmasked-tvla}
\end{figure}

\subsection{Resistance to Timing Attacks} 

Under the fixed-versus-random procedure of Section~\ref{sec:methodology}, all four timing statistics remained below the $|t|=4.5$ reference threshold without sustained growth (Fig.~\ref{fig:dudect-four-tests}). In the order shown in the figure legend, namely complete CCA2 decryption, complete CCA2 encryption, decryption polynomial multiplication, and encryption polynomial multiplication, the maximum running $|t|$ values were $1.589$, $1.037$, $1.634$, and $0.000$, respectively, while the final $|t|$ values were $0.578$, $0.126$, $0.405$, and $0.000$, respectively. Therefore, no class-dependent timing was detected for this $(256,256)$ binary in this experiment. This result provides empirical evidence rather than a universal proof of constant-time execution.

\subsection{Power-Trace Leakage}

Under the component-level procedure of Section~\ref{sec:methodology}, first-order TVLA was performed in successive stages of 5,000 traces, with multiple iterations used to confirm the observed behavior. The unmasked multiplication crossed the $|t|=4.5$ threshold, whereas the shared multiplication and the other protected operations remained below it (Table~\ref{tab:tvla_first_order} and Fig.~\ref{fig:masked-unmasked-tvla}). Masked multiplication and masked decoding, which test the protected core in isolation, stay well clear of the threshold; masked CCA2 decryption and CCA2 encryption sit closer to it because both traces also capture the unmasked CCA2 hashing, re-encryption, and comparison logic surrounding the protected core. Overall, no statistically significant leakage was detected in the evaluated operations under the applied test conditions.

\begin{table}[t]
\centering
\caption{First-order TVLA results for the protected operations.}
\label{tab:tvla_first_order}
\footnotesize
\setlength{\tabcolsep}{5pt}
\begin{tabular}{lc}
\hline
Operation & Max. $|t|$ \\
\hline
Masked multiplication    & 3.566 \\
Masked decoding          & 3.320 \\
Masked CCA2 decryption   & 4.360 \\
CCA2 encryption          & 4.194 \\
\hline
\end{tabular}
\end{table}

\subsection{Resistance to Fault-Injection Attacks}

The following results use the single-intervention logical-fault procedure of Section~\ref{sec:fault-campaigns}; Table~\ref{tab:fi-summary} summarizes 2,678 trials on the hardened builds. Preliminary testing had exposed two weaknesses: zeroing one comparison accumulator bypassed authentication, and zeroing the complete secret admitted an all-zero key in the selectable unshared-key build. Redundant comparisons and two-mask selection removed the former; structural validation and key erasure removed the latter in that configuration.

\begin{table}[t]
\centering
\caption{Logical fault-injection campaign summary.}
\label{tab:fi-summary}
\scriptsize
\renewcommand{\arraystretch}{1.08}
\begin{tabularx}{\columnwidth}{@{}l X r X@{}}
\toprule
\textbf{Fault} & \textbf{Injection target} & \textbf{Trials} & \textbf{Observed outcome} \\
\midrule
Zeroing & Selection, verification, and decryption state & 27 & Safe rejection \\
Zeroing & Complete secret in unshared-key build & 3 & Explicit key-generation error; key erased \\
Random & Forward/reverse selection masks & 512 & Safe rejection; both masks protected release \\
Random & Forward/reverse comparison accumulators & 2\,000 & Safe rejection; independent check remained \\
Random & Recovered decryption intermediate & 100 & Safe rejection by re-encryption check \\
Skip & Comparisons, mask generation, and mask use & 18 & Safe rejection through redundant path \\
Skip & Selector store and loop control & 6 & Silent corruption of valid output \\
Skip & Encryption randomness and tag generation & 6 & Explicit error returned \\
Skip & Sampling and validation in unshared-key build & 6 & Explicit error; invalid key rejected \\
\midrule
\textbf{Total} & \textbf{27 targeted location/value groups} & \textbf{2\,678} & \textbf{No authentication bypass or weak-key output} \\
\bottomrule
\end{tabularx}
\vspace{2pt}
\begin{minipage}{\columnwidth}
\raggedright
\footnotesize
\emph{Note:} The six silent corruptions caused neither authentication bypass nor weak-key output. Tests used debugger-assisted single-point logical faults at addresses resolved from the evaluated ELF.
\end{minipage}
\end{table}

All 2,612 value-substitution trials safely rejected, and none produced authentication bypass or weak-key output. The six selector-store and loop-control skip trials reveal a narrower integrity limitation: skipping the final store or loop control silently corrupts a valid output, so output-selection duplication or a loop-completion check remains necessary. No authentication bypass, candidate release, weak key, crash, or hang was observed after hardening. Zeroing-campaign records did not independently log each pre-injection value, so an already-zero target cannot be distinguished retrospectively from an effective zeroing fault. The conclusions remain limited to the tested single-point logical model.

\subsection{Comparison with Prior Embedded Lattice Implementations}

Table~\ref{tab:prior_performance_comparison} compares our branch-free CCA2/fault-protected unmasked configuration with prior embedded Binary Ring-LWE implementations. The comparison excludes our shared-key generation and first-order shared CPA decryption because the cited implementations do not report equivalent masking; schemes with substantially different arithmetic or
constructions are omitted.

\begin{table*}[t]
\centering
\scriptsize
\setlength{\tabcolsep}{2.5pt}
\renewcommand{\arraystretch}{0.95}
\caption{Embedded Binary Ring-LWE performance comparison.}
\label{tab:prior_performance_comparison}
\begin{adjustbox}{max width=\textwidth}
\begin{tabular}{@{} p{3cm} p{3.5cm} p{4cm} l l l @{}}
\toprule
Scheme / Implementation
& Platform
& Variant / Security Level
& Key Gen [Kcycles (ms)]
& Encrypt./Encaps. [Kcycles (ms)]
& Decrypt./Decaps. [Kcycles (ms)] \\
\midrule
PQ-IoTCrypt \cite{ma2024lightweight}
& ARM Cortex-M4F (168 MHz)
& IoT-Basic ($n$=256, $q$=2053)
& 7\,022.400 (41.8 ms)
& 6\,417.600 (38.2 ms)
& 5\,812.800 (34.6 ms) \\
PQ-IoTCrypt \cite{ma2024lightweight}
& ARM Cortex-M4F (168 MHz)
& IoT-Advanced ($n$=512, $q$=4093)
& 26\,241.600 (156.2 ms)
& 24\,141.600 (143.7 ms)
& 21\,806.400 (129.8 ms) \\
\midrule
Fault-resilient BRLWE \cite{ebrahimi2020lightweight}
& AVR ATxmega128A1
& $n$=256, $q$=256
& 1\,335.000 (39.8 ms$^{\dagger}$)
& 2\,691.000 (80.2 ms)
& 4\,037.000 (120.3 ms) \\
Fault-resilient BRLWE \cite{ebrahimi2020lightweight}
& ARM Cortex-M0
& $n$=256, $q$=256
& 1\,481.000 (44.1 ms$^{\dagger}$)
& 1\,755.000 (52.3 ms)
& 3\,404.000 (101.4 ms) \\
\midrule
\textbf{This work (Set 1)}
& \textbf{ARM Cortex-M33 (39 MHz)}
& \textbf{$n$=256, $q$=256}
& \textbf{330.053 (8.46 ms)}
& \textbf{604.186 (15.49 ms)}
& \textbf{871.055 (22.33 ms)} \\
\textbf{This work (Set 2)}
& \textbf{ARM Cortex-M33 (39 MHz)}
& \textbf{$n$=256, $q$=128}
& \textbf{332.078 (8.51 ms)}
& \textbf{609.302 (15.62 ms)}
& \textbf{879.130 (22.54 ms)} \\
\textbf{This work (Set 3)}
& \textbf{ARM Cortex-M33 (39 MHz)}
& \textbf{$n$=512, $q$=256}
& \textbf{1\,105.646 (28.35 ms)}
& \textbf{2\,065.859 (52.97 ms)}
& \textbf{3\,043.904 (78.05 ms)} \\
\bottomrule
\end{tabular}
\end{adjustbox}
\vspace{2pt}
\begin{minipage}{\textwidth}
\raggedright
\footnotesize
\emph{Note:} Times use native clocks; $^{\dagger}$ KeyGen times are inferred from reported cycle/time pairs. This work's three rows report cold-start, batched-TRNG results for the branch-free + CCA2/fault configuration, without shared decryption. Platform, parameter, and protection differences preclude direct ranking.
\end{minipage}
\end{table*}

\begin{figure}[t]
\centering
{\fontfamily{DejaVuSans-TLF}\selectfont
\begin{tikzpicture}[font=\scriptsize]
\def\radR{1.80}

\foreach \s in {0.25,0.50,0.75,1.00}{
  \draw[gray!28]
    (90:{\radR*\s}) -- (30:{\radR*\s}) --
    (-30:{\radR*\s}) -- (-90:{\radR*\s}) --
    (-150:{\radR*\s}) -- (150:{\radR*\s}) -- cycle;
}
\foreach \a in {90,30,-30,-90,-150,150}
  \draw[gray!45] (0,0) -- (\a:\radR);
\foreach \s in {0.25,0.50,0.75,1.00}
  \node[gray!65,anchor=west,font=\scriptsize] at (90:{\radR*\s}) {\s};

\node[align=center,anchor=south,font=\scriptsize] at (90:{\radR+0.18})
  {Encryption\\efficiency};
\node[align=left,anchor=west,font=\scriptsize] at (30:{\radR+0.18})
  {Decryption\\efficiency};
\node[align=left,anchor=west,font=\scriptsize] at (-30:{\radR+0.18})
  {CCA2-mechanism\\evidence};
\node[anchor=west,xshift=2pt,font=\scriptsize] at (-90:{\radR+0.10})
  {Fault protection};
\node[align=right,anchor=east,font=\scriptsize] at (-150:{\radR+0.18})
  {Constant-time\\evidence};
\node[align=right,anchor=east,font=\scriptsize] at (150:{\radR+0.18})
  {Power-SCA\\masking evidence};

\newcommand{\radarline}[7][]{%
  \draw[#1]
    (90:{\radR*#2}) -- (30:{\radR*#3}) --
    (-30:{\radR*#4}) -- (-90:{\radR*#5}) --
    (-150:{\radR*#6}) -- (150:{\radR*#7}) -- cycle;
}

\radarline[orange!85!black,very thick,densely dotted]
  {0.094}{0.586}{0}{0.5}{0.5}{0.5}
\radarline[red!75!black,very thick,dash dot]
  {0.345}{1}{1}{1}{0.5}{0}
\radarline[blue!75!black,very thick]
  {1}{0.857}{1}{1}{1}{1}

\begin{scope}[shift={(0.90,2.38)},font=\scriptsize]
  \draw[blue!75!black,very thick] (0,0) -- (0.45,0)
    node[right,black] {This work: Set~1};
  \draw[red!75!black,very thick,dash dot] (0,-0.34) -- (0.45,-0.34)
    node[right,black] {\cite{ebrahimi2020lightweight}: Cortex-M0};
  \draw[orange!85!black,very thick,densely dotted] (0,-0.68) -- (0.45,-0.68)
    node[right,black] {\cite{ma2024lightweight}: IoT-Basic};
\end{scope}
\end{tikzpicture}
}
\caption{Normalized encryption/decryption cycle efficiency and documented security-feature evidence. The comparison uses the final protected Set~1 configuration for this work, the $n=256$, $q=256$ Cortex-M0 result from~\cite{ebrahimi2020lightweight}, and the $n=256$, $q=2053$ IoT-Basic result from~\cite{ma2024lightweight}. Higher values are favorable, but the mixed platforms, moduli, and protection boundaries make the profiles descriptive rather than directly rank-equivalent.}
\label{fig:efficiency_security_radar}
\end{figure}
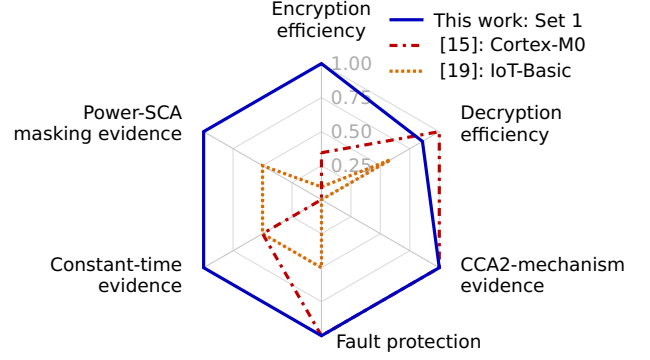

Because the implementations do not expose a common benchmark boundary, Fig.~\ref{fig:efficiency_security_radar} gives a normalized profile rather than a direct speedup claim. The chart's six axes are scored by two separate rules. The two efficiency axes (encryption, decryption) use $E_{i,o} = \min_j C_{j,o}/C_{i,o}$ for operation $o \in \{\mathrm{Enc}, \mathrm{Dec}\}$, where $C_{j,o}$ is the reported cycle count of implementation $j$ on operation $o$ and the minimum is taken over all compared implementations. The remaining four axes (CCA2-mechanism, fault protection, constant-time, and power-SCA masking evidence) use an ordinal rubric: 1 denotes an implemented feature supported experimentally or formally, 0.5 a claim unvalidated at the reported boundary, and 0 a feature not demonstrated in the paper; zero is not evidence of absence.

Our trace uses the final, fully protected Set~1 configuration, including masked decryption, so that efficiency and feature evidence refer to the same boundary; this differs from Table~\ref{tab:prior_performance_comparison}, which retains the unmasked CCA2/fault boundary because the cited works report no equivalent masked variant. Since encryption is unaffected by masking, our final configuration leads encryption efficiency outright; for decryption, our masked result is compared against the Cortex-M0 implementation of Ebrahimi and Bayat-Sarmadi~\cite{ebrahimi2020lightweight}, which leads on this axis since it carries no masking overhead. Ma et al.~\cite{ma2024lightweight} claim constant-time arithmetic, lightweight masking, and basic fault checks, but their Table~2 protection boundary is unclear and they do not demonstrate CCA2 security; those coordinates are therefore scored as claimed-only evidence.

\subsection{Discussion}
\label{sec:discussion}

The staged data show that byte packing, four-lane arithmetic, compact codecs, and buffered randomness accelerate RBLWE, whereas branch-free execution, CCA2-oriented verification, and sharing impose distinct costs. Cross-platform numbers remain indicative because schemes, parameters, security notions, and protected boundaries differ; our claim is a measurable transition from an accelerated CPA core to a hardened MCU implementation, not dominance over all prior work.
The reported security and DFR estimates apply only to the original uniformly sampled binary parameters and bias-aware decoder, not automatically to our custom, non-ML-KEM-compatible PKE wrapper. Sharing covers CPA decryption but not its surrounding oracle and re-encryption logic; fault results cover the tested single-point logical model. Future work should extend the evaluation through independent benchmark replication, higher-resolution power measurements, broader leakage tests, physical and multi-fault campaigns, and formal wrapper analysis for deployment-appropriate parameters.

\section{Conclusion}
\label{sec:conclusion}

This paper mapped RBLWE onto a commodity Cortex-M33 MCU using byte-lane SIMD arithmetic, a packed message codec, and a buffered Secure Element TRNG, yielding same-core speedups of $4.06\times$, $3.35\times$, and $3.01\times$ for key generation, encryption, and decryption. On top of this accelerated core, we staged four countermeasures, constant-time execution, fault hardening, a CCA2-oriented FO-style wrapper, and first-order shared CPA decryption, reporting each layer's cost individually. Binary-level verification confirmed these countermeasures survive compilation and caught a compiler-induced masking flaw, fixed in hand-written assembly. Dudect timing tests, component-level TVLA, and 2,678 fault trials found no exploitable leakage in the tested configurations, but surfaced one residual gap: an unduplicated output-selection step that a targeted instruction-skip fault can still corrupt, fixable by duplicating the store or adding a loop-completion check. The result is a measurable, staged transition from an accelerated CPA core to an implementation-hardened realization, with evidence and limitations reported at every stage.

\section*{Artifact Availability}
Source code, build scripts, binary-audit material, and evaluation data are available at \href{https://github.com/MuhammadAsimJavaid/rbinlwe-efm32pg28}{\texttt{github.com/MuhammadAsimJavaid/\allowbreak rbinlwe-efm32pg28}}.

\section*{Acknowledgment}
The authors used Claude (Anthropic) to generate the initial scaffolding of Figs.~2, 4, and 8, Algorithms~1--3, and Tables~I, II, and VI. All technical content and data are the authors' own or drawn from cited work; the AI system generated no results, claims, or scientific content. Claude was also used for grammar and prose-flow editing.

  \bibliographystyle{IEEEtran}
  \bibliography{references}

\end{document}